\documentclass[11pt]{amsart}
\newtheorem{theorem}{Theorem}[section]
\newtheorem{proposition}{Proposition}[section]

\newtheorem{lemma}{Lemma}[section]

\thanks{math. sub. classification: 81U05  and 47F05}
\begin{document}

\title{ Absolutely continuous  spectrum of a typical Schr\"odinger  operator with a  slowly decaying  potential}

\author{Oleg  Safronov }

\maketitle

\section{Main results}

We study  the absolutely continuous spectrum of a Schr\"odinger operator
\begin{equation}\label{op}
H=-\Delta+\alpha V,\qquad \alpha\in {\Bbb R},
\end{equation}
acting in the space $L^2({\Bbb R}^d)$. While the potential $V$ involved in this definition is a function of $x\in{\Bbb R}^d$, 
we shall also  study the dependence of $V$ on the spherical  coordinates $r=|x|$ and $\theta=x/|x|$. 
Therefore, sometimes the value of $V$ at $x\in {\Bbb R}^d$ will be  denoted by  $V(r,\theta)$. Even less often the radial variable will be denoted by $\rho.$
 Our main result is the following
\begin{theorem}\label{2} 
Let $V$ be a real valued bounded potential on ${\Bbb R}^d$ and let 
\begin{equation}\label{W}
W(  r,\theta)=\int_0^{ r} V( \rho,\theta)d \rho,\qquad  \forall r>0.
\end{equation}
Assume that $W$ belongs to the space ${\mathcal H}^1_{\rm loc}({\Bbb R}^d)$ of  functions having  (generalized)  locally square integrable derivatives. 
Suppose that
\begin{equation}\label{c1}
\int_{{\Bbb R}^d} \frac{|\nabla W|^2}{|x|^{d-1}}dx<\infty.
\end{equation}
Then the absolutely continuous spectrum of the operator \eqref{op} is essentially  supported by the interval $[0,\infty)$ for almost every $\alpha\in {\Bbb R}$. That is,
the spectral projection $E(\Omega)$ corresponding to any set $\Omega\subset[0,\infty)$ is different from zero $E(\Omega)\neq0$ as  soon as  the  Lebesgue measure of $\Omega$
is positive.
\end{theorem}

Note that in $d=1$, condition \eqref{c1} turns into 
\begin{equation}\label{DK}
\int_{{\Bbb R}}{|V|^2}dx<\infty.
\end{equation}
Operators with such potentials were studied in the work of Deift and Killip  \cite{DK}, the main result of  
which states that absolutely continuous spectrum of the operator $-d^2/dx^2+V$ covers the  positive half-line $[0,\infty)$,  if $V$ satisfies \eqref{DK}.

However,  it is  not clear 
what is  the proper generalization of  condition \eqref{DK} in $d\geq2$. Most likely,  it  should   be replaced by (cf. \cite{Simon})
\begin{equation}\label{simon}
\int_{{\Bbb R}^d} \frac{V^2}{|x|^{d-1}}dx<\infty,
\end{equation}
but the problem of proving  the existence of the absolutely continuous  spectrum under this  assumption turns out to be very hard.
Usually,  one assumes more than \eqref{simon}.
\newpage
For instance,  the result of the article \cite{Perelm} by Galina Perelman  says that
the absolutely continuous spectrum of the Schr\"odinger operator $$-\Delta+V$$
is essentially supported by $[0,\infty)$,  if
\begin{equation}\label{Perelm}
|V(x)|+|\nabla_\theta V(x)|\leq \frac C{(|x|+1)^{1/2+\epsilon}},\qquad \epsilon>0.
\end{equation}
Here, the symbol $\nabla_\theta V$ denotes the vector of derivatives with respect to the angular variables. A more precise definition of $\nabla_\theta$   is :
$$
\nabla=\frac xr \frac{\partial}{\partial r}+\frac1r {\nabla_\theta}.
$$
We see that, besides the decay of $V$ at infinity, the result of \cite{Perelm} requires a decay of $\nabla_\theta V$. Theorem ~\ref{2} has a similar assumption, however
it  deals with a wider  class of  potentials compared to the one  considered in \cite{Perelm}. Indeed,
 \eqref{Perelm} implies \eqref{c1}, but  the converse is not true.

The class  of functions described by the condition \eqref{c1} is wider  than the space of  functions satisfying \eqref{Perelm} 
not  only because the function under the integral sign in \eqref{c1} is allowed to have different behavior along  different directions, 
but also because the definition of $W$ involves  some averaging. For instance, the potential 
$$
V(r,\theta)=\frac1{r^{1/2+\delta} }\Bigl(2+\sin (r^{\gamma}\theta)\sin^n(\theta/2)\Bigr),\qquad  d=2,\,\, \theta\in [0,2\pi),
$$
fulfills the condition \eqref{c1} but does not satisfy \eqref{Perelm} if $\gamma>\delta$.

\section{Auxiliary material}

{\it Notations.} Throughout the text, ${\rm  Re}\, z$ and ${\rm Im}\,z$ denote the real and imaginary parts of a complex number $z$. 
The notation ${\Bbb S}$ stands  for the unit sphere in ${\Bbb R}^d$.  Its area is denoted  by $|{\Bbb S}|$.
For a selfadjoint operator $B=B^*$ and a vector $g$ of a Hilbert space the expression $((B-k-i0)^{-1}g,g)$
 is always understood as the limit
$$
\Bigl((B-k-i0)^{-1}g,g\Bigr)=\lim_{\varepsilon \to 0} \Bigl((B-k-i\varepsilon )^{-1}g,g\Bigr),\qquad   \varepsilon >0,\,\, k\in {\Bbb R}.
$$
This limit exists for almost every  $k\in {\Bbb R}$.

\bigskip

The following simple and very well known  statement plays very important  role in our proof.
\begin{lemma}\label{l1} Let $B$ be a self-adjoint operator in a separable Hilbert space ${\frak H}$ and let
$g\in{\frak H}$. Then the function
$$
\eta(k):={\rm Im}\,\Bigl((B-k-i0)^{-1}g,g\Bigr)\geq 0
$$
is integrable over ${\Bbb R}$. Moreover,
\begin{equation}\label{lemma}
\int_{-\infty}^\infty \frac{\eta(k)dk}{1+k^2}\leq \pi \Bigl((B^2+I)^{-1}g,g\Bigr).
\end{equation}
\end{lemma}

\bigskip

We  will also need  the following consequence  of  Hardy's inequality:
\begin{lemma}\label{l2} Let $V$ be a real valued potential vanishing inside the unit  ball and let $W$ be the function defined in \eqref{W}. 
Then
$$
\int_{|x|<R}\frac{ |W|^2}{|x|^{d+1}}dx\leq 4\int _{|x|<R}\frac{ |\nabla W|^2}{|x|^{d-1}}dx,
$$
\bigskip
for any $R>1.$
\end{lemma}

In the beginning of the proof of Theorem~\ref{2} we will assume that $V$ is compactly supported. 
We will obtain certain estimates on the derivative of the 
spectral measure of the operator $-\Delta+\alpha V$ for compactly supported potentials and then 
we will extend these estimates to the case of an arbitrary 
$V$ satisfying the conditions of Theorem~\ref{2}. 
We will approximate $V$ by  compactly supported functions. It is important not to destroy the  inequalities obtained previously for "nice" $V$.
Therefore the way we select approximations plays a very important  role in our proof.

 Let us describe our  choice of compactly  supported functions  $V_n$ approximating  the given potential $V$.
 Let us choose  a spherically symmetric function $\zeta\in {\mathcal H}^1({\Bbb R}^d)$ such that
 $$
 \zeta(x)=\begin{cases}
 1,\qquad {\rm if}\quad |x|<1;\\
 0,\qquad {\rm if}\quad |x|>2.
 \end{cases}
 $$
Assume  for simplicity that $0\leq \zeta\leq1$ and $|\nabla \zeta|\leq1$.
 Define $$\zeta_n(x)=\zeta(x/n).$$
Note that $\nabla \zeta_n\neq 0$ only in the shperical layer $\{\, x:\,\,n\leq |x|\leq 2n\}.$ Moreover $|\nabla \zeta_n|\leq 1/n$, which leads to the estimate $$|\nabla\zeta_n(x)|\leq 2/|x|.$$
 Our approximations of $V$ will be the functions $V_n$ defined as
 \begin{equation}\label{Qn}
 V_n=\frac \partial{\partial r}\, (\zeta_n W),
 \end{equation}
 where $W$ is the function from \eqref{W}. 
Thus,  approximations of $V$ by $V_n$ correspond to approximations of $W$ by
 \begin{equation}\label{apQ}
W_n=\zeta_n W.
\end{equation}
Observe  that, in this case,
$$
\int_{{\Bbb R}^d}\frac {|\nabla W_n(x)|^2}{|x|^{d-1}}dx\leq
\int_{|x|<2n}\frac {2|\nabla W(x)|^2+8|x|^{-2}|W(x)|^2}{|x|^{d-1}}dx \leq 34 \int_{{\Bbb R}^d}\frac {|\nabla W(x)|^2}{|x|^{d-1}}dx.
$$
Therefore,
\begin{equation}\label{c3}
\sup_n\int_{{\Bbb R}^d}\frac {|\nabla W_n(x)|^2}{|x|^{d-1}}dx<\infty.
\end{equation}
One can also easily show that
\begin{equation}\label{Vinfty}
||V_n||_\infty\leq  3||V||_\infty.
\end{equation}

\section{Proof of Theorem~\ref{2}}

Our proof is based on the relation between the
derivative of the spectral measure and the so called  scattering amplitude. Both  objects should be
introduced properly. While the spectral measure can be defined  for any  linear self-adjoint mapping,
the scattering  amplitude  will be introduced only for a differential operator.
Let $f$ be a vector in the Hilbert space ${\frak H}$ and $H$ be a self-adjoint operator in ${\frak H}$.
It turns out that the quadratic form of the resolvent of $H$ can be written as a Cauchy integral
$$
((H-z)^{-1}f,f)=\int_{-\infty}^\infty\frac{d\mu(t)}{t-z},\qquad {{\rm Im}\,z}\neq0.
$$
 The measure $\mu$ in this representation is called the spectral measure of $H$ corresponding to the element $f$.

Let us now  introduce the scattering amplitude. Assume that
the support of  the potential $V$  is compact
and take any compactly supported function  function  $f$. Then
$$
(H-z)^{-1}f=e^{ik|x|}\frac{ A_f(k,\theta)}{|x|^{(d-1)/2} }+O(|x|^{-(d+\delta)/2}),
\qquad  {\rm as}\,\, |x|\to\infty,\,\,\theta=\frac{x}{|x|}, \,\, k^2=z,\,\, {\rm Im}\,k\geq0,\,\,\delta>0,
$$ with some $A_f(k,\theta)$.
Clearly, the relation
$$
\mu'(\lambda)=\pi^{-1}\lim_{z\to\lambda+i0}{\rm Im}\,((H-z)^{-1}f,f)=\pi^{-1}\lim_{z\to\lambda+i0}{\rm Im}\,z ||(H-z)^{-1}f||^2
$$
implies that
\begin{equation}\label{pimu}
\pi\mu'(\lambda)= \sqrt\lambda\int_{\Bbb S} |A_f(k,\theta)|^2\,d\theta, \qquad  k^2=\lambda>0.
\end{equation}
Formula \eqref{pimu} is a very important estimate that
relates the absolutely continuous  spectrum to so-called extended  states. The rest of the proof will be devoted to a lower estimate of $|A_f(k,\theta)|$.

Consider  first  the case $d=3$.
For our purposes, it is sufficient  to assume that   $f$ is the characteristic function of the unit ball.  In this case, $f$ is a spherically symmetric function.
Traditionally, $H$ is viewed as an operator  obtained by a perturbation of   $$H_0=-\Delta.$$
In its turn,  $( H-z)^{-1}$ can be viewed as an operator  obtained by a perturbation of $( H_0-z)^{-1}$.
The  theory of such perturbations is often based on  the second resolvent identity
\begin{equation}\label{hilbert}
( H-z)^{-1}=( H_0-z)^{-1}-(H-z)^{-1}\alpha V( H_0-z)^{-1},
\end{equation}
which turns  out to be useful for  our reasoning.
As a  consequence of \eqref{hilbert}, we obtain that
\begin{equation}\label{ak2}
A_f(k,\theta)=F(k)+A_{g}(k,\theta),\qquad z=k^2+i0,\,\,k>0,
\end{equation}
where $g (x)=\alpha V(x)( H_0-z)^{-1}f$ and $F(k)$ is defined  by
\begin{equation}\label{rea}
(H_0-z)^{-1}f=e^{ik|x|}\frac{ F(k)}{|x|^{(d-1)/2} },
\qquad  {\rm for}\,\, |x|>1 \qquad ({\rm recall\,\,\,\, that }\,\,d=3).
\end{equation}
 Without loss  of generality, one can assume that $V(x)=0$ inside the unit ball. In this case,
\begin{equation}
\label{psi=}
g=F(k) h_k,\qquad {\rm where}
\quad h_k(x)=
\alpha V(x)e^{ik|x|}|x|^{(1-d)/2}.
\end{equation}
According to \eqref{pimu},  \eqref{ak2} and  \eqref{psi=},  we obtain 
\begin{equation}\label{triterms}
\pi \mu'(\lambda)\geq |F(k)|^2 \Bigl(|{\Bbb S}|\sqrt\lambda - {\rm Im}\,\Bigl( ( H-z)^{-1} h_k,\, h_k\Bigr)\Bigr).\end{equation}
Therefore, in order to establish the presence of the absolutely continuous spectrum, we  need to show that the quantity
$
{\rm Im}\,\Bigl( ( H-z)^{-1} h_k,\, h_k\Bigr)
$  is small. The chain of the arguments that led us to this conclusion has been suggested by Boris Vainberg.
  The  method  developed   by  the author in the previous  version of the paper  was  much  longer.

Let us define $\eta_0$ setting
$$
\alpha^2k^{-2} \eta_0(k,\alpha):=\frac1{k}{\rm Im}\,\Bigl( ({H}-z)^{-1} h_k,\, h_k\Bigr)\geq0.$$ Obviously $\eta_0$ 
is positive for all real $k\neq 0$, because  we agreed  that $z=k^2\pm i0$  if $\pm k>0.$
This is very convenient.
Since $\eta_0>0$, we can  conclude that $\eta_0$ is small on a rather large set if the integral of this function is small. 
That is why we will try to estimate
\begin{equation}\label{J}
J(V):=
\int_{-\infty}^\infty \int_{-\infty}^{\infty}
\frac{\eta_0(k, \alpha)}{(\alpha^2+k^2)}\,\frac{|k|\, dkd\alpha}{(k^2+1)}=\int_{-\infty}^\infty \int_{-\infty}^{\infty}\frac{\eta_0(k, tk)}{(k^2+1)(t^2+1)}\,dkdt.
\end{equation}
Now,  we employ a couple of  tricks, one of which has an artificial character and will be appreciated not immediately but a bit later.  
Instead of dealing  with the operator ${H}$, we will deal with ${H}+\varepsilon I$ where $\varepsilon>0$ is a small parameter.
We will first  obtain an integral estimate for    the quantity
$$
\eta_\varepsilon(k,\alpha)=\frac{k}{\alpha^2}{\rm Im}\,\Bigl( ({H}+\varepsilon -z)^{-1} h_k,\, h_k\Bigr).
$$
The latter estimate will be not uniform in $\varepsilon$, but we can still pass to the limit $\varepsilon\to0$ according  to Fatou's lemma,  because
$$
\eta_0(k,\alpha)=\lim_{\varepsilon\to0}\eta_\varepsilon(k,\alpha) \qquad {\rm  a.e.\,\,on}\,\, 
{\Bbb  R}\times{\Bbb  R}.
$$

The second  trick is to set $\alpha=kt$ and represent $\eta_\varepsilon$ in the form
\begin{equation}\label{eta}
\eta_\varepsilon(k,kt)={\rm Im}\,\Bigl( (B+1/k)^{-1} H_\varepsilon^{-1/2} v,\, H_\varepsilon^{-1/2}v\Bigr)
\end{equation}
where $v=V|x|^{(1-d)/2}$, $H_\varepsilon=-\Delta+\varepsilon I$ and $B$ is the bounded selfadjoint operator defined by
$$
B=H_\varepsilon^{-1/2}\Bigl(-2i\frac{\partial}{\partial r}-\frac{i(d-1)}{|x|}+tV\Bigr)H_\varepsilon^{-1/2}.
$$
The symbol $r$ in the latter formula denotes   the radial variable $r=|x|$. 
The reader can easily establish that $B$ is not only self-adjoint but bounded as well. Note that  it is the parameter $\varepsilon$ that makes $B$ bounded.

In order to justify \eqref{eta} at least formally, one has to introduce 
the operator $U$ of multiplication by the function $\exp(ik|x|)$. Using  this notation, we can represent $\eta_\varepsilon$ in the following  form
$$
\eta_\varepsilon(k,tk)=k{\rm Im}\,\Bigl(U^{-1} ({H}+\varepsilon -z)^{-1}U v,\, v\Bigr).
$$
Since  we deal with a unitary equivalence of operators, we can employ the formula
$$
U^{-1} ({H}+\varepsilon -z)^{-1}U= (U^{-1}{H}U+\varepsilon -z)^{-1}.
$$
On the other hand, since ${H}$ is a differential operator and $U$ is an operator of multiplication, the commutator $[{H},U]:={H}U-U{H}$ can be easily found
$$
\Bigl[{H},U
\Bigr]=kU\Bigl(-2i\frac{\partial}{\partial r}-\frac{i(d-1)}{|x|}+k\Bigr).$$
The latter equality implies that
$$
U^{-1}{H}U+\varepsilon-z=H_\varepsilon+k\Bigl(-2i\frac{\partial}{\partial r}-\frac{i(d-1)}{|x|}+tV\Bigr)=H_\varepsilon^{1/2}(I+kB)H_\varepsilon^{1/2}.
$$
Consequently,
\begin{equation}\label{apendix}
k U^{-1} ({H}+\varepsilon -z)^{-1}U=H_\varepsilon^{-1/2}(B+1/k)^{-1}H_\varepsilon^{-1/2}.
\end{equation}
 A more detailed  proof of \eqref{apendix} will be given in the last section called "Appendix".
These details do not have  so much value  for us at the moment. It is more important that,  now,  \eqref{eta} follows  from \eqref{apendix}.

Let us have a look at the formula \eqref{eta}. If $k$ belongs to the upper half plane then so does $ -1/k$.
Since $B$ is a self-adjoint operator, $\pi^{-1}\eta_\varepsilon(k,kt)$ coincides with the derivative of the spectral measure of the operator $B$ corresponding to the element $H_\varepsilon^{-1/2}v$.
According to Lemma~\ref{l1}, the latter observation implies that
$$
\int_{-\infty}^\infty \frac{ \eta_\varepsilon(k,kt)}{\,(1+k^{2}) } dk\leq \pi \Bigl( (B^2+I)^{-1}H_\varepsilon^{-1/2}v, H_\varepsilon^{-1/2}v\Bigr),
$$
which leads to 
\begin{equation}\label{B^{-2}}
\int_{-\infty}^\infty \frac{ \eta_\varepsilon(k,kt)}{\,(1+k^{2}) } dk\leq \pi \Bigl( B^{-1}H_\varepsilon^{-1/2}v,  B^{-1}H_\varepsilon^{-1/2}v\Bigr)=\pi || B^{-1}H_\varepsilon^{-1/2}v||^2.
\end{equation}
Our further  arguments  will be related to the estimate of the quantity in the right hand side of \eqref{B^{-2}}.
We will   show now that
\begin{equation}\label{show}
\lim_{\varepsilon\to0}|| B^{-1}H_\varepsilon^{-1/2}v||^2\leq C \int_{{\Bbb R}^d} \frac{|\nabla W|^2}{|x|^{d-1}}dx.
\end{equation}
In order to do that we use the representation 
 \begin{equation}\label{TAT}
B^{-1}H_\varepsilon^{-1/2}=H_\varepsilon^{1/2} T^{-1},
\end{equation}
where $T\subset T^*$ is the first order differential  operator, defined by 
$$
T=-2i\frac{\partial}{\partial r}-\frac{i(d-1)}{|x|}+tV,\qquad D(T)=D(H_\varepsilon^{1/2}).
$$
The representation \eqref{TAT} is a simple consequence of  the fact that
$
B=H_\varepsilon^{-1/2}TH_\varepsilon^{-1/2}.
$

Let us discuss the basic properties of the operator $T$. The study of these properties is rather simple,   
because one can derive an  explicit  formula for the resolvent of $T$.
For that purpose, one needs to recall the  theory of ordinary differential  equations, which says that
the equation
$$
y'+p(t)y=f(t),\qquad y=y(t),\,\, t\in {\Bbb R},
$$ is equivalent to the relation
$$
\Bigl(e^{\int p\,dt}y\Bigr)'=e^{\int p\,dt}f.
$$
Put  differently,$$
y'+p(t)y=e^{-\int p\,dt}\Bigl(e^{\int p\,dt}y\Bigr)'.
$$
This  gives  us a clear  idea of  how to handle the operator $T$.
 Let $U_0$ and $U_1$ be the operators of multiplication
by
$|x|^{(d-1)/2}$ and by $\exp( 2^{-1}it W)$, then
$$
T=-2iU_1^{-1}U_0^{-1}\Bigl[\frac{\partial}{\partial r}\Bigr]U_0U_1,\qquad {\rm and}\qquad T^{-1}=\frac i2U_1^{-1}U_0^{-1}\Bigl[\frac{\partial}{\partial r}\Bigr]^{-1}U_0U_1.
$$
Since $[\frac{\partial}{\partial r}]^{-1}$ means just the   simple integration with respect to $r$ and $\partial W/\partial r=V$,
\begin{equation}\label{Tg}\begin{split}
T^{-1}v=\frac i2 e^{-2^{-1}it W}|x|^{-(d-1)/2}\int_0^r e^{2^{-1}it W} V dr=\\
\frac{1}t e^{-2^{-1}it W}|x|^{-(d-1)/2}(e^{2^{-1}it W} -1)=\frac{1}t |x|^{-(d-1)/2}(1-e^{-2^{-1}it W} ).\end{split}
\end{equation}
Note, that $T^{-1}v$ turns out to be compactly supported, which leaves no doubt about the relation $v\in D(T^{-1})$.
Combining \eqref{TAT} with \eqref{Tg}, we conclude that
$$
\lim_{\varepsilon\to0}|| B^{-1}H_\varepsilon^{-1/2}v||^2\leq ||\nabla T^{-1}v||^2\leq C\Bigl(\int_{{\Bbb R}^d} \frac{| W|^2}{|x|^{d+1}}dx+ 
\int_{{\Bbb R}^d} \frac{|\nabla W|^2}{|x|^{d-1}}dx\Bigr).
$$
 Now \eqref{show} follows  from Lemma ~\ref{l2}.
We remind the reader that   \eqref{B^{-2}}, \eqref{show} are needed to estimate
the quantity $J(V)$ from \eqref{J}.   We can say  now that
$$
J(V)\leq C\int_{{\Bbb R}^d} \frac{|\nabla W|^2}{|x|^{d-1}}dx.
$$
Using  Chebyshev's  inequality, we derive  from  \eqref{triterms} that 
\begin{equation}\label{chebysh}
\begin{split}
{\rm meas}\,\,\{   (\lambda,\alpha)\in   [\lambda_1,  \lambda_2]\times [\alpha_1,\alpha_2]:\,\,\,\,\,\,\,\,|{\Bbb S}|\sqrt\lambda-\frac{\pi \mu'(\lambda)}{ |F(k)|^2}>s\}
\leq 
C s^{-1}
\int_{{\Bbb R}^d} \frac{|\nabla W|^2}{|x|^{d-1}}dx
\end{split}
\end{equation}
for finite $\lambda_j>0$ and $\alpha_j>0.$ The constant $C>0$ in \eqref{chebysh} depends on $||V||_\infty$,  $\lambda_j>0$ and $\alpha_j>0.$ 
If  $s=2^{-1}|{\Bbb S}|\sqrt\lambda_1$ then \eqref{chebysh} turns into
\begin{equation}\label{chebysh2}
\begin{split}
{\rm meas}\,\,\{   (\lambda,\alpha)\in   [\lambda_1,  \lambda_2]\times [\alpha_1,\alpha_2]:\,\,\,\,\,\,\,\,
\frac{\pi \mu'(\lambda)}{ |F(k)|^2}<|{\Bbb S}|(\sqrt\lambda -2^{-1}\sqrt\lambda_1)\}
\leq 
C _0
\int_{{\Bbb R}^d} \frac{|\nabla W|^2}{|x|^{d-1}}dx.
\end{split}
\end{equation}

We can say  now that the proof is  more or less completed,
because the quantity in the right hand side 
can be made  arbitrary small if  we replace $V$ by $V-V_n$, where $V_n$ are defined in \eqref{Qn} and $n$ is sufficiently large. Put  differently, we  keep the   "tails"  of $V$ and remove only a compactly  supported portion of it.
The latter operation changes $V$ only on a compact set. According to the Scattering Theory, this operation does not change the absolutely continuous spectrum of the
 Schr\"odinger operator  $-\Delta+V$. This implies, that without loss of generality, we can assume that  $\pi\mu'(\lambda)\geq 2^{-1}{|{\Bbb S}|\sqrt\lambda}{|F(\sqrt \lambda)|^2}$on a  set of  a large measure.

\section{Semi-continuity of the entropy}

Let us  complete the proof and mention the missing ingredients. First, in order to understand what we achieved, we summarize the results. We found such approximations of $V$ by compactly supported  potentials $V_n$ that
the corresponding spectral measures $\mu_n$ satisfy the estimate (see \eqref{chebysh2}):
\begin{equation}\label{mu>f}
\pi\mu'_n(\lambda)\geq \frac{|{\Bbb S}|\sqrt\lambda}2{|F(\sqrt \lambda)|^2}= \frac{|{\Bbb S}|}{2\lambda^{5/2}}
\Bigr|\sqrt \lambda\cos(\sqrt \lambda )-\sin(\sqrt \lambda )\Bigl|^2
\end{equation}
on a set of pairs $(\lambda,\alpha)$ of very large Lebesgue measure.
Denote the characteristic function of the intersection of this set with the rectangle $[\lambda_1,\lambda_2]\times[\alpha_1,\alpha_2]$  by $\chi_n$. 
Thus, inequality \eqref{mu>f} holds on the support of $\chi_n.$ (By the way, we assume that $\lambda_1>0$  and $\alpha_1>0$ are positive.)

We will study the behavior of $\chi_n$ as $n\to\infty$.
The difficulty of the situation is that $\chi_n$ might  change with the growth of $n$. However,
 since the unit ball in any Hilbert space is compact in the weak topology,
without loss of generality, we can assume that $\chi_n$ converges weakly in $L^2$ to a square integrable function $\chi$. 
In a certain  sense, we can say that  $\chi_n$  does not change much if $n$ is  sufficiently large. Now the situation is less  hopeless, 
because the limit $\chi$ preserves  properties of the sequence $\chi_n.$
 The necessary  information about the limit $\chi$ can be easily obtained from the information about $\chi_n$.
 It is clear that $0\leq \chi\leq1$ and $\chi>0$ on a set of very large  measure $(\lambda_2-\lambda_1)(\alpha_2-\alpha_1)-\varepsilon$. Indeed, let $\tilde\chi$ be the characteristic function of the set where $\chi>1+\varepsilon_0$.
 Since $\int\int \chi_n \tilde{\chi} d \lambda\,d\alpha\leq \int\int\tilde{\chi}  d \lambda\,d\alpha$,
 we obtain that
 $$
 (1+\varepsilon_0 ) \int\int \tilde{\chi} d \lambda\,d\alpha\leq\int\int \tilde{\chi}  d \lambda\,d\alpha,
 $$
 which is possible only in the case  when $\tilde\chi=0$ almost everywhere. Consequently, $\chi\leq1$
 and therefore we can judge about the size of the set where $\chi>0$ by the value of the integral $\int\int \chi \,d \lambda\,d\alpha=\lim_{n\to\infty}\int\int \chi_n \,d \lambda\,d\alpha$.

 It is also known, that if $V_n$ converges to $V\in L^{\infty}({\Bbb R}^d)$ in $L^2_{loc}$, then
\begin{equation}\label{mu-n}
\mu_n\to\mu \qquad {\rm as}\,\, n\to \infty
\end{equation}
weakly for any fixed $\alpha$. We see that both sequences $\mu_n$ and $\chi_n$ have a limit,
however they converge in a weak sense, which brings additional difficulties.
Therefore we have to find a quantity  that  not only  depends on a pair of measures  (semi-)continuously with respect to the weak topology,
 but  is also infinite as soon as the derivative of one of the measures $\mu'=0$ vanishes on a large set. Such a quantity is the entropy, defined by
$$
S=\int_{\alpha_1}^{\alpha_2}\int_{\lambda_1}^{\lambda_2} \log\Bigl(\frac{\mu'(\lambda)}{\chi(\lambda,\alpha)}\Bigr)\chi(\lambda,\alpha)\,d\lambda d\alpha.
$$
Its properties were thoroughly studied in \cite{KS}. It can diverge only to negative infinity, but if it is finite, then $\mu'>0$ almost  everywhere on the set  $\{(\lambda,\alpha): \,\,\chi>0\}$.
We can formulate a more general definition:

\bigskip
{\it Definition.}  Let $\rho,\, \nu$ be finite Borel measures on a compact Hausdorff
space, $X$. We define the entropy of $\rho$ relative to $\nu$  by
\begin{equation}\label{semi}
 S( \rho | \nu) =\begin{cases}
-\infty, \qquad {\rm if}\,\,  \rho\,\,{\rm  is\,\, not} \,\,\,\nu\,{\rm -ac}\\
-\int_X \log(
\frac{ d\rho}{d\nu })d\rho,\qquad {\rm  if}\,\,\, \rho\,\,{\rm  is} \,\,\nu\,{\rm-ac}.\end{cases}
\end{equation}

\bigskip

\begin{theorem}\label{semiKS}{\rm (cf.\cite{KS})} The entropy 
$ S( \rho | \nu)$ is  jointly upper semi-continuous in $\rho$ and $\nu$ with respect to the  weak topology. That is,
if $\rho_n\to\rho$ and $\nu_n\to \nu$ as $n\to\infty$, then
$$
S(\rho|\nu)\geq \limsup_{n\to\infty}S(\rho_n|\nu_n).
$$

\end{theorem}

Relation \eqref{mu-n} literally  means that
$$
\int \phi(\lambda,\alpha)\,d\mu_n\to \int \phi(\lambda,\alpha)\,d\mu\qquad {\rm as}\,\, n\to \infty,
$$
for any fixed $\alpha$ and any continuous compactly supported function $\phi$. By the Lebesgue dominated convergence theorem, we obtain that
$$
\int \int \phi(\lambda,\alpha)\,d\mu_n d\alpha\to \int \int \phi(\lambda,\alpha)\,d\mu d\alpha\qquad {\rm as}\,\, n\to \infty,
$$
which means that the sequence of measures 
$$
{\rm meas}_n\, (\Omega ):=\int \int_{(\lambda,\alpha)\in \Omega} d\mu_n\,d\alpha
$$
converges weakly as well. Now, Theorem~\ref{semiKS}  implies that
\begin{equation*}
\int_{\alpha_1}^{\alpha_2}\int_{\lambda_1}^{\lambda_2} \log\Bigl(\frac{\mu'(\lambda)}{\chi(\lambda,\alpha)}\Bigr)\chi(\lambda,\alpha)\,d\lambda d\alpha\geq \liminf_{n\to\infty}\int_{\alpha_1}^{\alpha_2}\int_{\lambda_1}^{\lambda_2} \log\Bigl(\frac{\mu_n'(\lambda)}{\chi_n(\lambda,\alpha)}\Bigr)\chi_n(\lambda,\alpha)\,d\lambda d\alpha>-\infty,
\end{equation*}
because  logarithmic integrals are semi-continuous with respect to weak convergence of measures.
This proves that $\mu'>0$ on the support of $\chi$  which is a subset of $[\lambda_1,\lambda_2]\times[\alpha_1,\alpha_2]$ whose measure is not smaller than  $(\lambda_2-\lambda_1)(\alpha_2-\alpha_1)-\varepsilon$. It remains to observe that $\varepsilon$ is arbitrary.

This  proves our main result for $d=3$. Now, if $d\neq 3$, then  equality  of the form \eqref{rea} is  incorrect. We  have to deal with the terms of smaller order that 
must  appear in the right hand side of \eqref{rea}. However,  one can  avoid this difficulty  replacing the operator
$H_0$ by $$
H_0=-\Delta-\frac{\kappa_d  \tilde \chi}{|x|^2}P_0,\qquad \kappa_d=\Bigl(\frac{d-2}{2}\Bigr)^2-\frac14, 
$$
where $P_0$ is the projection onto the space of  spherically symmetric  functions and  $\tilde \chi$ is the characteristic function of the compliment of the unit ball (cf. \cite{summer11}). 
$$
{}
$$

The  semi-continuity of logarithmic integrals \eqref{semi} was discovered for the broader audience  by R.Killip and B.Simon in \cite{KS}. The reason why $S$ is semi-continuous is that $S$ is representable as
an infimum of a difference of two integrals with respect to the measures $\nu$ and $\rho$:
$$
S(\rho |\nu ) =\inf_F\Bigl(
\int
F(x) d\nu -
\int
(1 +\log F(x)) d\rho\Bigr),\qquad \min_x F(x)>0.
$$

In conclusion of this section, we would like to draw your attention to the   papers
 \cite{D1}-\cite{LNS}, \cite{Perelm}-\cite{summer11} which contain an important work 
on the absolutely continuous spectrum of multi-dimensional Schr\"odinger operators. Two of these papers (\cite{D}, \cite{summer11}) deal with  families of Schr\"odinger operators
$-\Delta+\alpha V$, where $V$ is not only decaying but oscillating as well.

\section{Appendix}

Here we prove  the relation \eqref{apendix}. 
If $k$ is not real, then $U^{-1}$ is an  unbounded operator. However, this fact does not bring additional difficulties, because we  will apply the operator $U^{-1}$ only  to  functions that  decay at infinity sufficiently fast.
Let us   formulate now the  statement which justifies \eqref{apendix}.
\begin{proposition} Let  $V$ be a compactly  supported real potential. 
 Let $k$ be a point in the upper half-plane, let $z=k^2$ and
let $v \in L^2({\Bbb R}^d)$   
be a  compactly supported function. Then 
$$
u=k ({H}+\varepsilon -z)^{-1}U v,\qquad \alpha=kt,\,\,\,\, {\rm Im}\,z\neq 0.
$$
is  representable  in the form 
$$
u=Uw
$$
with 
$
w\in  L^2({\Bbb R}^d).
$
Moreover,
$$
w=H_\varepsilon^{-1/2}(B+1/k)^{-1}H_\varepsilon^{-1/2}v,
$$
where $H_\varepsilon=-\Delta+\varepsilon$ and
\begin{equation}\label{Bdef}H_\varepsilon^{1/2}BH_\varepsilon^{1/2}v=-2i\frac{\partial v}{\partial r}-\frac{i(d-1)v}{|x|}+tV v .
\end{equation}
\end{proposition}

\bigskip

{\it Proof.}   Consider the function  $w=e^{-ik|x|}u$. It is easy to see that $w$ is a solution of the differential equation
$$
-\Delta w +\varepsilon w+ktVw-2ik\frac{\partial w}{\partial r}-\frac{ik(d-1)w}{|x|}=k \, v.
$$
Moreover,
 $w$ decays at infinity as $O\Bigl(e^{-({\rm Im}\,\sqrt{k^2-\varepsilon}-{\rm Im}\,k)|x|}\Bigr)$. Consequently, 
$w\in D(H_\varepsilon)$ and 
$$
H_\varepsilon w+k H_\varepsilon^{1/2}BH_\varepsilon^{1/2} w=k\,v
$$
The proof  is completed.  $\,\,\,\,\,\,\,\,\,\,\,\, \Box$

\bigskip

Another statement which might help the reader to understand our arguments, deals with analytic properties of the resolvent of $H$.
\begin{proposition}   Assume that $V$ is a bounded compactly supported  potential.
Let $\chi$ be the characteristic  function of a compact set containing the support of $V$.
Then the operator valued  function
$$
T(k)=\chi(H-k^2)^{-1}\chi,\qquad \alpha=kt,
$$
admits a  meromorphic continuation into  the  plane with the cut along  the half-line $\{\, z=i y,\,\,y\leq0\,\}$. 
\end{proposition}

\bigskip

{\it Proof.} Indeed, if $V=0$, then the  proof of the statement  can be found in \cite{Vainberg}.
Note that
$$
T_0(k)=\chi(H_0-k^2)^{-1}\chi,\qquad k\in {\Bbb C}\setminus\{\, z=i y,\,\,y\leq0\,\} ,
$$
is an integral operator whose kernel  depends on $k$ analytically. Moreover, the results of \cite{Vainberg} clearly say  that $T_0(k)$ is compact.
The relation  $$\chi(H-k^2)^{-1}\chi=\chi(H_0-k^2)^{-1}\chi-tk\chi(H_0-k^2)^{-1}V\chi (H-k^2)^{-1}\chi$$
implies that
$$
T(k)=\Bigl(I+tkT_0(k)V\Bigr)^{-1}T_0(k).
$$
The statement follows now from the analytic Fredholm alternative. $\,\,\,\,\,\,\,\,\,\,\,\, \Box$

\end{document}